\documentclass{jpsj-suppl}
\usepackage{txfonts} 

\def\vect#1{{\mbox{\boldmath $#1$}}}

\def\ketv#1{\mid \hspace{-0.5mm} #1 \rangle}
\def\brav#1{\langle #1 \hspace{-1mm} \mid}

\title{Pairing Effects in Nuclear Fusion Reaction}

\author{Shuichiro \textsc{Ebata}$^{1}$ and Takashi \textsc{Nakatsukasa}$^{2}$}

\inst{$^{1}$Meme Media Laboratory, Hokkaido University, Sapporo-shi 060-0810, Japan\\
$^{2}$RIKEN Nishina Center, Wako-shi 351-0198, Japan}

\email{ebata@nucl.sci.hokudai.ac.jp}

\recdate{July 15, 2013}

\abst{
We simulate a heavy-ion collision using the canonical-basis time-dependent 
Hartree-Fock-Bogoliubov theory (Cb-TDHFB) treating pairing correlation 
in the three-dimensional coordinate space. 
We apply the Cb-TDHFB to $^{22}$O+$^{22}$O collision with a contact-type pairing energy functional, 
and compare results of Cb-TDHFB and TDHF to investigate the effects of pairing correlations in nuclear fusion. 
Our results seem to indicate that pairing effects do not increase the fusion cross section in this system.
}

\kword{Heavy-ion collision, Pairing, Time-dependent mean field theory}

\begin{document}
\maketitle
\section{Introduction}
The time-dependent Hartree-Fock theory (TDHF) is well-known as a useful tool to study nuclear dynamics\cite{KD77}. 
However, TDHF can not describe effects of pairing correlation which has basically attractive property in the ground state 
and plays an important role in nuclear structure and low-energy nuclear reactions. 
There is a theory to treat pairing correlation in nuclear dynamics self-consistently, 
which is called the time-dependent Hartree-Fock-Bogoliubov theory (TDHFB). 
So far, no study of heavy-ion collision has been done using the TDHFB with a modern effective 
interaction in three-dimensional space, 
because of a number of numerical difficulties and requirement of the huge computational resource\cite{BL10}. 

In order to study nuclear dynamics treating pairing correlation, 
we proposed a new time-dependent mean-field theory 
which is named the canonical-basis TDHFB (Cb-TDHFB)\cite{EN10}. 
The Cb-TDHFB is derived from full TDHFB equations represented in the canonical basis 
which diagonalizes the density matrix, using the BCS-like 
approximation for pairing functional ($pp$,$hh$-channel). 
We confirmed the validity of Cb-TDHFB for the linear response calculations, 
comparing the results with those of the quasi-particle random phase approximation (QRPA) 
which is a small amplitude limit of the full TDHFB\cite{EN10}.

In the present work, we apply the Cb-TDHFB to a large amplitude collective motion, 
namely the heavy-ion collision. 
The simulation has been done in the three-dimensional Cartesian coordinate space 
using the Skyrme effective interaction (SkM$^{\ast}$) with the contact pairing. 
In Sec.2, we introduce the Cb-TDHFB equations and the pairing energy functional on the present work. 
Then, in Sec.3, we show the difference between fusion reactions 
with and without pairing correlation in $^{22}$O+$^{22}$O. 

\section{Formulation}
\subsection{Cb-TDHFB equations and contact pairing}
The Cb-TDHFB equations can be derived from the TDHFB equations 
with an approximation for pairing functional\cite{EN10}. 
The TDHFB state at any time  can be expressed in the canonical (BCS) form. 
\begin{eqnarray}
|\Phi (t) \rangle \equiv 
\prod_{l>0} \Big( u_{l}(t) + v_{l}(t) \hat{c}_{l}^{\dag}(t)\hat{c}_{\bar l}^{\dag}(t)  \Big) | 0 \rangle, 
\label{eq:BCS}
\end{eqnarray}
where $u_{l}(t),v_{l}(t)$ are time-dependent BCS factors and 
$\{ \hat{c}_{l}^{\dag}, \hat{c}_{\bar l}^{\dag}\}$ are creation operators of canonical pair of states. 
This is guaranteed by the Bloch-Messiah theorem\cite{RS80}. 
Then, we choose the BCS form of pair potential as
\begin{eqnarray}
\Delta_{l}(t) = - \sum_{k>0} \kappa_{k}(t)\ \bar{\cal V}_{l\bar{l},k\bar{k}}(t) \ ,
\label{eq:delta}
\end{eqnarray}
where $\kappa_{k}(t) \equiv u_{k}(t)v_{k}(t)$ 
corresponds to the pairing-tensor $\kappa (t)$ in the canonical-basis 
and $\bar{\cal V}_{l\bar{l},k\bar{k}}$ is the anti-symmetric two-body matrix element 
that has time dependence due to the time-evolution of canonical-basis. 
The subscripts $\bar{l}$ and $\bar{k}$ mean canonical partners of $l$ and $k$, respectively. 
We can obtain the Cb-TDHFB equations with the approximate pair potential Eq.(\ref{eq:delta}), as follows. 
\begin{eqnarray}
i \hbar \frac{\partial \phi_{l}(t)}{\partial t} &=& (\hat{h}(t) - \eta_{l}(t))\ \phi_{l}(t), \ \ i \hbar \frac{\partial \phi_{\bar l}(t)}{\partial t}  = (\hat{h}(t) - \eta_{\bar{l}}(t))\ \phi_{\bar{l}}(t),  \nonumber \\
i \hbar \frac{\partial \rho_{l}(t)}{\partial t}  &=& \kappa_{l}(t)\Delta_{l}^{\ast}(t) - \kappa_{l}^{\ast}(t)\Delta_{l}(t),  \nonumber \\
i \hbar \frac{\partial \kappa_{l}(t)}{\partial t}  &=& (\eta_{l}(t) + \eta_{\bar l}(t) )\ \kappa_{l}(t) + \Delta_{l}(t) (2\rho_{l}(t) - 1), 
\label{eq:Cb-TDHFB}
\end{eqnarray} 
where $\eta_{l}(t) \equiv \brav{\phi_{l}(t)} \hat{h}(t)\! \ketv{\phi_{l}(t)} + i \hbar \brav{\frac{\partial \phi_{l}}{\partial t}}\! \phi_{l}(t) \rangle$. 
The Cb-TDHFB equations are composed of the canonical basis $\phi_{l}(t), \phi_{\bar l}(t)$, 
the occupation probability $\rho_{l}(t) \equiv |v_{l}(t)|^{2}$ and the pair probability $\kappa_{l}(t)$. 
They conserve the orthonormal property of the canonical basis and the average particle number. 
When we choose a special gauge condition $\eta_{l}(t) = \varepsilon_{l}(t) = \brav{\phi_{l}(t)}\hat{h}(t) \! \ketv{\phi_{l}(t)}$, 
they conserve average total energy. 

We introduce neutron-neutron and proton-proton BCS pairing. 
The BCS pairing matrix elements ${\cal V}_{l\bar{l},k\bar{k}}^{\tau}$ is written as 
\begin{eqnarray}
 {\cal V}_{l\bar{l},k\bar{k}}^{\tau}=\int\! d\vect{r}_{1}d\vect{r}_{2} \sum_{\sigma_{1},\sigma_{2}} 
 \phi_{l}^{\ast}(\vect{r}_{1},\sigma_{1})\phi_{\bar{l}}^{\ast}(\vect{r}_{2},\sigma_{2})
 \hat{\cal V}^{\tau}(\vect{r}_{1},\sigma_{1};\vect{r}_{2},\sigma_{2}) \nonumber \\
 \times \left[  \phi_{k}(\vect{r}_{1},\sigma_{1})\phi_{\bar{k}}(\vect{r}_{2},\sigma_{2})
 -\phi_{\bar{k}}(\vect{r}_{1},\sigma_{1})\phi_{k}(\vect{r}_{2},\sigma_{2}) \right]. \hspace{-15mm}
\label{eq:Viijj0}
\end{eqnarray}
We introduce the spin-singlet contact interaction to Eq.(\ref{eq:Viijj0}): 
\begin{eqnarray}
\hat{\cal V}^{\tau}(\vect{r}_{1},\sigma_{1};\vect{r}_{2},\sigma_{2}) 
\equiv V_{0}^{\tau} \frac{1-\hat{\vect{\sigma}}_{1}\cdot \hat{\vect{\sigma}}_{2}}{4} \delta(\vect{r}_{1}-\vect{r}_{2}), 
\label{eq:Vint}
\end{eqnarray}
where $\tau$ indicates neutron or proton channel and $V_{0}^{\tau}$ is a strength of pairing\cite{KB90}. 

\subsection{Numerical details}
Before the time-evolution, 
we prepare the ground states of projectile and target nuclei 
performing the self-consistent HF+BCS calculation. 
The initial state of the simulation is constructed by locating the wave functions of projectile and target 
at an impact parameter $b$ and at a relative distance $H$ where they interact through only the Coulomb interaction. 
We boost the wave functions and calculate 
the time-evolution of nuclear densities following Eq.(\ref{eq:Cb-TDHFB}). 
In the present work, we use the three-dimensional Cartesian coordinate-space representation for the 
canonical states, $\phi_{l}(\vect{r},\sigma; t) = \brav {\vect{r},\sigma}\! \phi_{l}(t) \rangle$ 
with $\sigma=\pm 1/2$. 
These wave functions of $^{22}$O are calculated in a 20-fm cube. 
The coordinate space for collision dynamics is a rectangular box of 32 fm $\times$ 20 fm $\times$ 40 fm, 
discretized in the square mesh of $\Delta x = \Delta y = \Delta z = 1.0$ fm. 

\section{Results} 
We simulate the $^{22}$O+$^{22}$O collision with an incident energy $E_{\rm cm}=10$ MeV 
which is higher than the Coulomb barrier of present system. 
Initial distance $H$ between projectile and target is 20 fm along z-axis and 
the impact parameter is $b=$ 2.7 $\sim$ 3.1 fm along x-axis. 
$^{22}$O ($Z$=8, $N$=14) has superfluidity only for neutrons. 
The number of canonical basis treated in the calculation is 48 for the total system. 
The neutron pairing strength $V_{0}^{\rm n}$ is defined to reproduce the gap energy 
obtained from experimental binding energies using three-points formula.
The average neutron gap energy 
$\bar{\Delta}^{\rm n} \equiv \sum_{l>0} \Delta_{l}^{\rm n} / \sum_{l>0}$ is 2.06 MeV. 

\begin{figure}[h]
\begin{center}
\includegraphics[width=140mm, clip]{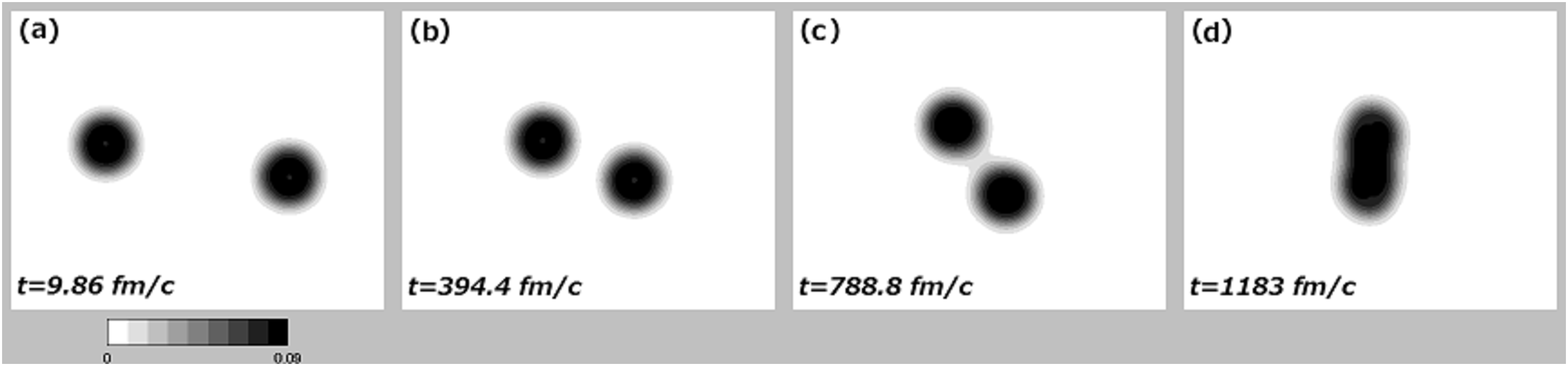}
\caption{
Neutron density distributions of $XZ$-plane in $^{22}$O+$^{22}$O collision 
at $t$=($a$)9.86, ($b$)394.4, ($c$)788.8 and ($d$)1183 ${\rm fm / c }$. 
They are results of TDHF simulation with $E_{\rm cm}$=10 and $b$=3 fm.}
\ \\[5mm]
\includegraphics[width=140mm, clip]{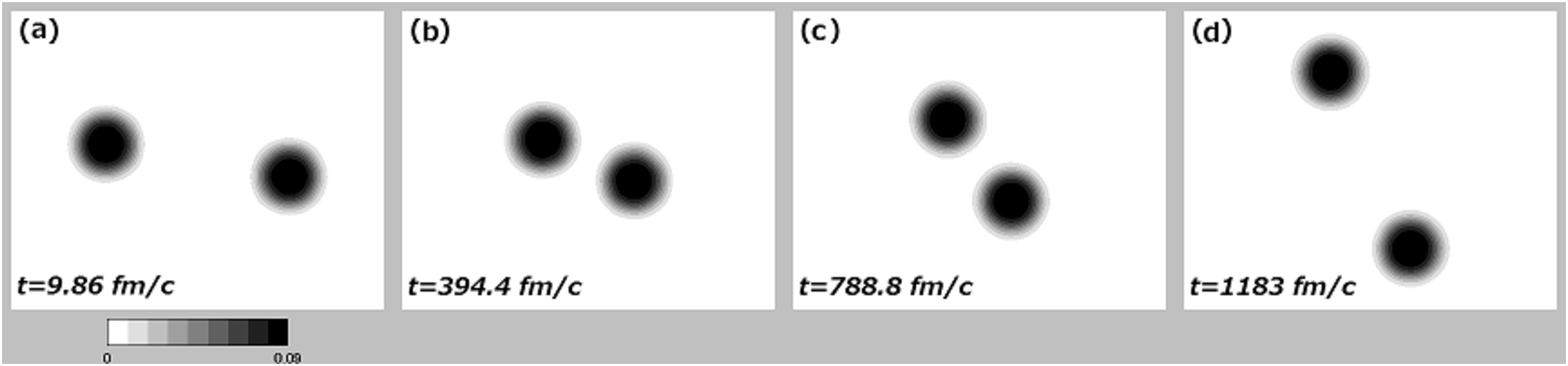}
\caption{
Same as Figure 1, but these results are obtained on Cb-TDHFB simulation.}
\end{center}
\end{figure}

Figure 1 and 2 indicate the time-evolution of neutron density distributions 
for the $^{22}$O+$^{22}$O collision phenomena for the simulation of TDHF and Cb-TDHFB, respectively, 
with a impact parameter $b$=3.0 fm. 
We can see a remarkable difference between the results with and without pairing correlation. 
Before two nuclei touch (panels (a) and (b)), 
the TDHF and Cb-TDHFB these simulation show almost same behavior. 
Near the touching point, (c) in Fig.1 and 2, the difference appears in 
thin neutron-density distribution. 
In the TDHF simulation, a neck-like distribution is formed, 
but they are separated completely in the Cb-TDHFB simulation.
After the touching: (d) panels, 
two nuclei fuse in the TDHF while they do not in the Cb-TDHFB calculation. 
We can evaluate a fusion cross section $\sigma_{F}$ using: 
\begin{eqnarray}
\sigma_{F} \equiv 2\pi \int_{0}^{b_{lim}} b\ db, 
\end{eqnarray}
where $b_{lim}$ is the maximum impact parameter to fuse. 
We can say, at least, that the fusion cross section $\sigma_{F}^{\rm B}$ obtained in the Cb-TDHFB simulation
is smaller than that of TDHF simulation $\sigma_{F}^{\rm H}$. 
We also check the pairing strength dependence of the fusion cross section. 
We choose a weaker pairing strength and simulate the fusion reaction with the same procedure. 
We have confirmed the relation that the fusion cross section $\sigma_{F}^{\rm Bw}$
 in the weak pairing strength is lager than $\sigma_{F}^{\rm B}$. 
These seem to indicate that the pairing correlation does not enhance the fusion in present system. 

We simulated the $^{22}$O+$^{22}$O collision using the time-dependent mean field theory including 
the effects of nuclear pairing, and have shown that the fusion properties are different between 
with and without pairing correlation. 
In the present system, the pairing correlation hinders the fusion cross section. 
We need further study, to obtain a definite calculation. 
Currently, we investigate the quenching effects in more heavy mass region 
and in the combination of different nuclei. 

\section*{Acknowledgement} 
The simulation has been supported by the high performance computing system 
at Research Center for Nuclear Physics, Osaka University, 
the RIKEN Integrated Cluster of Clusters(RICC) and the SR16000 at  YITP in Kyoto University.

\end{document}